\documentstyle[aps]{revtex}
\input epsf.sty
\begin{document}
\draft
\flushbottom
\twocolumn[
\hsize\textwidth\columnwidth\hsize\csname @twocolumnfalse\endcsname

\title{Andreev scattering in nanoscopic junctions at high 
magnetic fields.}
\author{H. Suderow$^1$,
E. Bascones$^2$, W. Belzig$^3$, F. Guinea$^2$ and S. Vieira$^1$}
\address{$^1$ Laboratorio de Bajas Temperaturas, 
Departamento de F{\'\i}sica de la Materia Condensada,
 Instituto de Ciencia de Materiales Nicol\'as Cabrera,
Facultad de Ciencias, C-III, Universidad Aut\'onoma de
Madrid. E-28049 Madrid. Spain \\$^2$ Instituto de Ciencia de Materiales de 
Madrid,
Consejo Superior de Investigaciones
Cient{\'\i}ficas,
Cantoblanco, E-28049 Madrid, Spain \\
$^3$Delft University of Technology, Laboratory of Applied Physics and DIMES,
2628 CJ Delft, The Netherlands.}
\date{\today}
\maketitle
\tightenlines
\widetext

\begin{abstract}
We report on the measurement of multiple Andreev resonances at atomic 
size point contacts between two superconducting nanostructures of Pb under 
magnetic fields higher than the bulk critical field, where superconductivity
is restricted to a mesoscopic region near the contact. The small number of 
conduction channels in this type of contacts permits a quantitative 
comparison with theory through the whole field range. 
We discuss in detail
the physical properties of our structure, in which the normal bulk electrodes
induce a proximity effect into the mesoscopic superconducting part.

\end{abstract}

]
\narrowtext
\tightenlines

It is well know that it is possible to fabricate atomic size contacs 
between metallic electrodes in a controlled way by means of the mechanically
 controllable break junction 
technique or the scanning 
tunneling microscope (STM).
\cite{MRJ92,A93}
Indeed, by repeatedly indenting the tip
into the sample one 
can achieve a stationary state  
in which a connecting neck between the electrodes is formed.\cite{AR93,UR97} 
This neck elongates
 and contracts during the repeated indentation following 
a well defined pattern of elastic and plastic steps, which has
 been neatly measured in a combined STM-AFM experiment where conductance and 
forces could be 
recorded simultaneously\cite{R93}.

The properties of a given neck can be probed by measuring the current-voltage
 characteristic within the same experiment, so that the STM serves at the 
same time as a fabrication tool and 
as an experimental probe of a very singular atomic size 
nanostructure.\cite{UR97,R93}
A reasonable knowledge of the geometry of the neck which can be 
varied in a well controlled way, is obtained through 
a
 simultaneous measurement of the
 conductance during the fabrication process\cite{UR97}.
  The final form of these structures, which are successfully 
fabricated \cite{UR97,P93,PB98} is a long connecting neck jointed on its ends to the bulk 
electrodes 
whose radius decreases in a smooth way towards a central constriction, which can 
be of atomic size.

In this experiment, the control of the morphology extends 
over two lengths scales: first, the overall form of the neck 
can be varied at mesoscopic length scales (hundreds or thousands of \AA) by the 
repeated indentation 
process, and second, the smallest cross section can be varied at 
atomic scales (tens of \AA) by doing small voltage variations on the z-
piezotube.
 Recently  new possibilities of atomic size contacts
  have lead to progress on the understanding of some phenomena 
occurring 
at a nanoscopic level. It has been shown during the last years that lead (Pb) is 
a good material
 to create this kind of small dimensions systems having the additional advantage 
of 
being a superconductor below $T_c=7.16K$.\cite{AR93} Indeed, the transport of current 
between two weakly linked 
superconductors brings noteworthy information about the contact through e.g. the 
Josephson 
current or through the multiple Andreev reflection mechanism.\cite{Ti}
 In the case of a single atom link between two 
electrodes the authors of Ref.\cite{S97} proposed that the effect of the 
multiple
 Andreev resonances on the I-V 
characteristics is a measure of the number and transparency of the conduction 
channels through 
a single atom\cite{S98}.

In this work we focus on the magnetic field dependence of the 
I-V characteristics of single atom point contacts.
Indeed, it is well known that superconductors of 
 reduced dimensions such as thin films or granular samples remain 
 superconducting well above Hc.\cite{Ti} As the magnetic field penetration depth of 
lead
 is about 390\AA  \ for a bulky sample, it is feasible to build connecting
  necks with smaller
 lateral dimensions with the repeated indentation 
 procedure.\cite{UR97,PB98}
We find indeed that sufficiently long and narrow necks show superconducting features 
up to fields as large as 20 times the bulk critical field of Pb (which is 0.05T at 1.5K).
The structure of subgap resonances due to multiple Andreev 
reflections remains under field and our analysis shows in detail how the pair 
breaking effect of the magnetic field, together with the 
N-S proximity effect from the bulk electrodes smears the subgap 
resonances.

We use a stable STM setup with a tip and a sample of the same 
material (Pb) which is brought from the tunneling into the contact regime 
by cutting the feedback loop. The I-V curves were taken at 1.5K using a standard four 
wire technique. Great care was taken to shield electrically the whole 
setup as RF noise is known to smear the subgap resonances in small contacts.
The experiment is done by gently changing the smallest cross section of the neck 
to make a large number of atomic size contacts at each 
magnetic field without varying the overall form of the neck.
Indeed, while our setup is sufficiently stable to maintain the same neck over 
a complete magnetic field sweep, we cannot maintain the morphology of the neck on the 
atomic level over a
 large field variation. Nevertheless, we could perform small field sweeps of 
several hundred Gauss with a given atomic arrangement and we find the same result,
 so that the measurement procedure does not change the results presented here.
The maximal elongation of the piezotube, 
which is 1600 \AA, limits the overall length of the necks.
Here we discuss one typical case of a neck having a critical field of about
 20 times the bulk critical field
 of lead with a magnetic field applied always parallel to the long 
axis. 

\begin{figure}
\epsfysize 12cm
\hspace{0cm}
\epsfbox{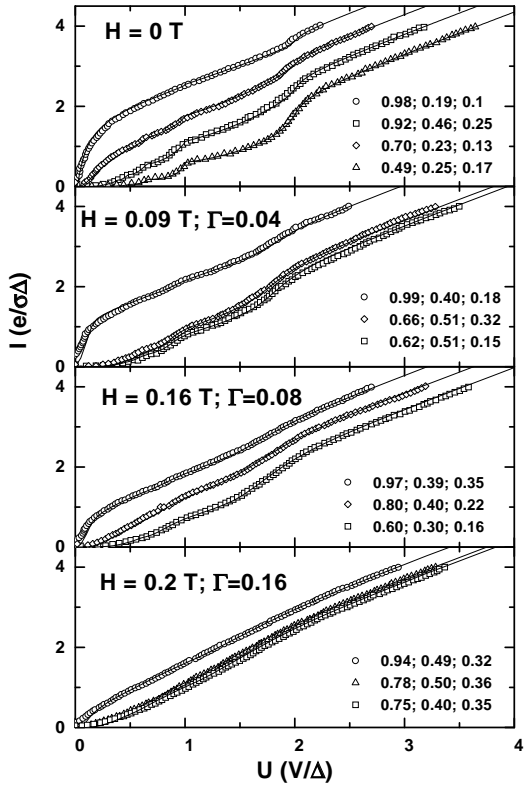}
\caption{
I-V characteristics of atomic sized contacts in the presence of
a magnetic field taken at 1.5K. $V$ is normalized to
 $\Delta$ at each field (from the top to the bottom:
$\Delta_0$, $0.89\Delta_0$, $0.78\Delta_0$, 
$0.7\Delta_0$ with the measured $\Delta_0=1.35meV$), and $I$ is also normalized to $\Delta$
 and to $\sigma$, the conductance above $2\Delta$ (as expressed in units 
of the quantum of conductance $\sigma_0=2e^2/h$). 
 Not all experimental points are plotted in order to show 
more clearly the fits (straight lines) 
using the model explained in the text.
 The parameters in the lower right corner
are the transmissions through the different
channels used to fit the experimental data (see text). The sum
 over the three parameters gives the total conductance $\sigma$ 
in units of $\sigma_0$. Each
line of numbers corresponds to one curve, from top to bottom.
$\Gamma$ is the pair breaking parameter (also defined in the text).
The upper critical field of bulk Pb is 0.05 T.}  
\label{fig:fittings}
\end{figure}

 Fig.1 shows a 
representative choice of measured I-V curves 
 of several last contacts 
before breaking 
at zero field and under field.
Each I-V curve, is different at each contact and can be well fitted
at zero field by the conduction 
channels model of Ref.\cite{S97} (straight lines, upper figure in Fig.1).
Accordingly, the experimental I-V curves in a last contact 
show a 
large variety of behaviors which is slightly changed by varying the 
morphology of 
the contact at atomic length scales.
This is compatible with measurements that record the 
conductance at a fixed voltage in a large number of contacts and show 
how the conductance of Pb last contacts shows steps
comparable to the quantum unit of conductance, 
but the average value has large fluctuations\cite{SirventTesis}.
The model of Ref.\cite{S97} 
uses (and verifies) the theoretical predictions that the I-V curve 
between two superconductors which are weakly linked through a small number of 
conduction channels is highly non linear and varies strongly 
depending on the 
transparency $T$ of the junction ($0<T<1$; tunnel to contact regimes). It turns out, that 
only one conduction channel with a given $T$ is not sufficient to fit the I-V curves shown in Fig.1, 
but that it is necessary to add a number $N$ of theoretical 
curves, each one with a given $T_n$ between 0 and 1.
This was related to the number $N$ and transparency $T_n$ of channels in each single atom 
contact, where $N$ and the average values of the $T_n$'s depend on the element studied.
In the case of Pb, this gives $N=3$ with $T_1$ rather opened (most frequently close to $1$) 
and $T_{2,3}$ more closed (smaller than $1$). In the
 uppermost part of Fig.1 the numbers show the experimentally measured $T_n$'s 
for each contact. We 
will 
not go into more details about this model which is extensively discussed in 
Refs.\cite{S97,S98,CML96,BG99}.
In the following, we discuss how to explain the data under field.

We first analyze the influence of the magnetic
field introducing the pair breaking effect in the standard procedure\cite{S97},
 as formulated in
a wavefunction representation\cite{BG99,ZA98,A85,AB95}. 
It was shown in\cite{ZA98}
that pair breaking effects can be incorporated by modifying the
Andreev reflection amplitude, 
$a ( \omega ) = u ( \omega ) - \sqrt{u^2 ( \omega ) - 1}$,
where $u ( \omega )$ satisfies\cite{M69}:
\begin{equation}
\frac{\omega}{\Delta} = u \left( 1 - \Gamma \frac{1}{\sqrt{1 - u^2}} \right)
\label{contact}
\end{equation}
where $\Gamma= 1 / 
( \Delta \tau_{pb} )$,
$\tau_{pb}$ is the pair breaking time 
and $\Delta$ is the self consistent superconducting
gap including the pair breaking effects. This expression is generally
valid, irrespective of the origin of the
pair breaking mechanism\cite{M69}.
The value of $\Gamma$ used in the fittings
was assumed to be the same
for all channels and all I-V curves at a given applied field.

The straight lines in Fig.1 shows the fittings which are as good as the 
ones obtained at zero field, provided
that the pair breaking parameter is introduced. The number $N$ and the 
characteristic values 
of the parameters for the transparency of each channel $T_n$
 does not vary up to the largest fields. $\Gamma$ is determined 
with a precision of about 20\%.

The values of $\Gamma$ explain 
the magnetic field dependence
 of the gap in the tunneling regime,
which we have measured by breaking completely the contact.
 We therefore could follow and fit precisely the 
predicted influence of pair breaking in the structures associated with multiple 
Andreev 
reflections. This was not possible to do previously, as other realizations of 
multiple
Andreev reflections (e.g. large point contacts, tunnel junctions with 
microbridges \cite{Ti}) involve experimental setups which are much more 
complex than a single atom 
contact with a small number of conduction channels and cannot be modelled 
precisely.

We can gain more insight in the physics of this system if we consider 
 the pair
breaking parameter in a uniform superconducting cylinder in a magnetic 
field which is given by \cite{M69}:
\begin{equation}
\frac{\hbar}{\tau_{pb}} = \frac{e^2 D R^2 H^2}{6 \hbar^2}
\label{pairbreaking}
\end{equation}
where $R$ is the radius of the cylinder, $D = v_F l / 3$ is the diffusion
coefficient, $l$ is the mean free path, and $H$ is the applied field.
Following this model, in order to explain the values of $\Gamma$
 used in fig.[\ref{fig:fittings}] we need a cylinder of a radius which is 
rather large ($R \approx 450$\AA, taking $\xi \approx R$, note that a smaller value 
of $\xi$ leads to even larger values of $R$)
 as compared to the usual width estimations 
for the neck presented here or other necks fabricated with the same method
\cite{UR97}. Clearly, a model based on a simple cylinder does not 
explain the observed behavior, we need to take into account that the radius varies as 
a function of $z$.

We analyze in the following the order parameter, density of states and 
pair breaking parameter in a neck of a varying radius.
Indeed, a better agreement is obtained if we consider that at a given 
field, the superconducting region is in good contact with the part of 
the neck with larger radius which already became normal, so that pair 
breaking effects arise from the proximity effect of this normal region.
Assuming that the electronic mean free path is smaller than the coherence 
length,
we can describe the superconducting properties 
by the Usadel equations\cite{U70,RS86,BW99}. We parametrize the Green's
functions in terms of an angle parameter, $\theta ( \vec{r} , E )$, where
$E$ is the energy measured from the chemical potential\cite{NS96}.
Setting $\hbar = 1$, they can be written as:
\begin{eqnarray}
\frac{D}{2} \nabla^2 \theta &+ & i E 
\sin ( \theta ) + | \Delta | \cos ( \theta ) 
\nonumber \\ & -  
&2 e^2 D | \vec{A} |^2  \cos ( \theta ) \sin
( \theta ) = 0
\label{usadel}
\end{eqnarray}
where 
$\vec{A} = ( H r \vec{u}_{\phi} ) / 2$ is the vector potential. 
We neglect the influence of other spin flip and inelastic
processes. At the boundary
of the contact, we have $| \nabla \theta |_{R} = 0$, and $R ( z )$ 
determines the geometry of the neck (we neglect any radial dependencies, 
and take $R(z) \le \xi$).
We also 
assume that the magnetic field is unscreened
within the neck.
Then, $A$ can be replaced by its average,
$\langle A^2 \rangle  ( z ) = \frac{H^2 R(z)^2}{12}$. 
Within this approximation, the vector potential enters
in eq.[\ref{usadel}] as giving rise to an effective, position dependent,
pair breaking time. If we apply eq. [\ref{usadel}] to
a uniform wire, this pair breaking time reduces to that in
eq.[\ref{pairbreaking}]. 

\begin{figure}
\epsfysize5.5 cm
\hspace{0cm}
\epsfbox{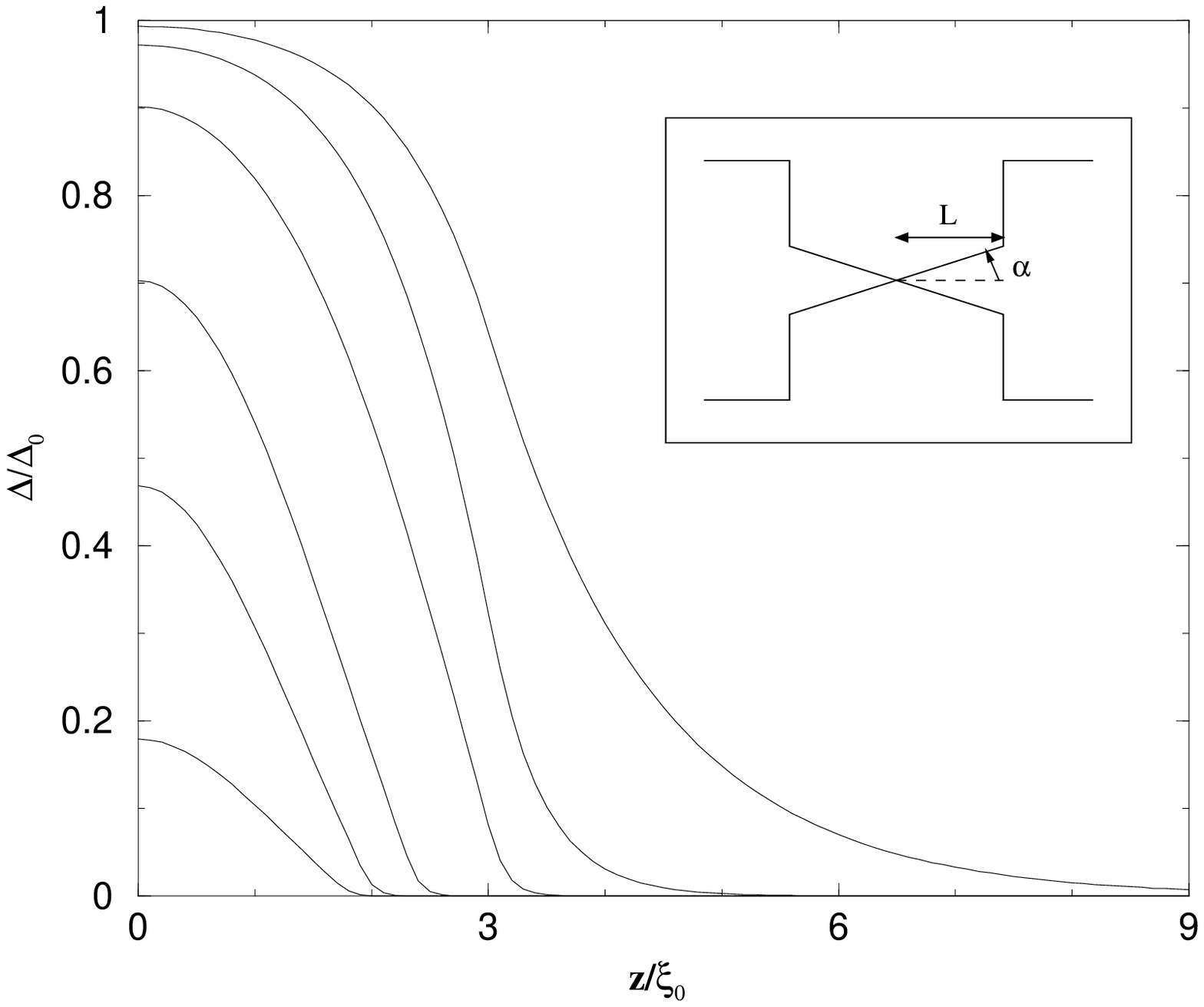}
\caption{
Superconducting order parameter as function of distance,
for different applied fields. The contact region is located
at $z / \xi = 0$ (from top to bottom, $H=0.1,0.2,0.6,1.0,1.2,1.3$ T). 
The magnetic field is applied along the long axis of the cones.} 
\label{fig:gap}
\end{figure}

\begin{figure}
\epsfysize5.5 cm
\hspace{0cm}
\epsfbox{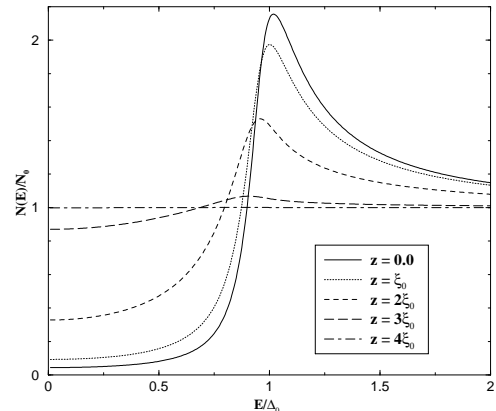}
\caption{
Electronic density of states for different positions
of the neck. The field is $H =0.2$T.}    
\label{fig:dosz}
\end{figure}

Figure [\ref{fig:gap}] shows the superconducting order parameter,
for different fields, as function of the position for a typical neck  
modelled by two truncated cones of $L = 800$ \AA \, length 
attached to bulk electrodes, with an 
opening angle of $\alpha = 35^0$.  We also take
$L = 3 \xi$, so that $\xi \approx 260 \AA $.\cite{Nota1}
There is a smooth
transition to the normal state as the radius of the neck increases.
This is further illustrated in figure[\ref{fig:dosz}], where the
density of states is shown at different positions
for $H = 0.2$T. For this field,
the influence of the normal region is felt throughout the 
entire neck. Even at the central region, the gap is significantly rounded.
This is also observed in the calculated density of states at
the center shown in Fig.4.

\begin{figure}
\epsfysize=5.5 cm
\hspace{0cm}
\epsfbox{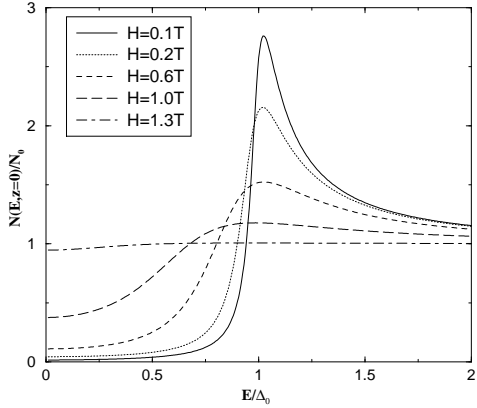}
\caption{
Density of states at the contact region for different 
applied fields.}
\label{fig:field}
\end{figure}

>From the solution of the Usadel equations we can infer the
amplitude for Andreev reflections at the contact surface,
which is given by $i \tan [ \theta ( z=0 , E ) / 2 ]$.
This function is slightly different from the standard
expression used to incorporate pair breaking effects in a
point contact (see equ. [\ref{contact}] and Ref.\cite{ZA98}). We have checked that
there are no appreciable differences in the quality of the
fits to the experimental data, shown in fig.[\ref{fig:fittings}], 
with a reasonable value for $\xi \approx 260 - 300 \AA$.

In conclusion, we have measured, and analyzed, the multiple 
Andreev scattering resonances of atomic sized Pb contacts in the presence 
of a magnetic field greater than the bulk critical field.
In this regime, superconductivity is restricted to a small
neck of mesoscopic dimensions.
We are able to build and control in situ with our STM 
structures
which are a unique example of weak links 
of dimensions variable from atomic to mesoscopic length 
scales, opening a new field of studies in nanophysics.
We present a quantitative comparison of experiment and theory of 
pair breaking effects on 
multiple Andreev resonances. 

We would like to thank discussions and the help of A. Izquierdo, 
G. Rubio and N. Agra{\"\i}t.
One of us (E. B.) is thankful to the Universit\"at Karlsruhe for
its hospitality. Financial
support from the TMR program of the European Commission under
contract ERBFMBICT972499, the CICyT (Spain) through grant PB96-0875, 
the CAM (Madrid) through grant 
07N/0045/98 and FPI and the spanish DGIGyT under contract PB97-0068 
are gratefully acknowledged.


\begin{references}
\bibitem{MRJ92}
C. J. Muller, J. M. van Ruitenbeek and L. J. de Jongh,
Phys. Rev. Lett. {\bf 69}, 140 (1992).
\bibitem{A93}
N. Agra{\"\i}t, J. G. Rodrigo and S. Vieira, Phys. Rev. B {\bf 47},
12345 (1993).
\bibitem{AR93}
N. Agra{\"\i}t, J. G. Rodrigo, C. Sirvent and S. Vieira, Phys. Rev. B {\bf 48},
8499 (1993).
\bibitem{UR97}
C. Untiedt, G. Rubio, S. Vieira and N. Agra{\"\i}t, Phys. Rev. B
{\bf 56}, 2154 (1997).
\bibitem{R93}
G. Rubio, N. Agra{\"\i}t, S. Vieira, Phys. Rev. Lett. {\bf 76}, 2032 (1996).
\bibitem{P93}
J. I. Pascual, J. M\'endez, J. G\'omez-Herrero, A. M. Bar\'o,
N. Garc{\'\i}a and V. Thien Binh, Phys. Rev. Lett. {\bf 71}, 1852 (1993).
\bibitem{PB98}
M. Poza, E. Bascones, J. G. Rodrigo, N. Agra{\"\i}t, S. Vieira and
F. Guinea, Phys. Rev. B {\bf 58}, 11173 (1998).
\bibitem{Ti}
M. Tinkham, "Introduction to Superconductivity", Second Edition, McGraw Hill (1996).
\bibitem{S97}
E. Scheer, P. Joyez, D. Esteve, C. Urbina and M. H. Devoret,
Phys. Rev. Lett. {\bf 78}, 3535 (1997).
\bibitem{S98}
E. Scheer, N. Agra{\"\i}t, A. Cuevas, A. Levy-Yeyati, B. Ludolph,
A. Mart{\'\i}n-Rodero, G. Rubio, J. M. van Ruitenberg and C. Urbina,
Nature {\bf 394}, 154 (1998).
\bibitem{SirventTesis} 
C. Sirvent, PhD thesis, Universidad Autonoma de Madrid.
\bibitem{CML96}
J. C. Cuevas, A, Mart{\'\i}n-Rodero and A. Levy-Yeyati. Phys. Rev. B
{\bf 54}, 7366 (1996).
\bibitem{BG99}
E. Bascones and F. Guinea, preprint.
\bibitem{ZA98}
A. V. Zaitsev and D. V. Averin, Phys. Rev. Lett. {\bf 80},
3602 (1998).
\bibitem{A85}
G.B. Arnold, J. Low Temp. Phys. {\bf 59}, 143 (1985).
\bibitem{AB95}
D. V. Averin and A. Bardas, Phys. Rev. Lett. {\bf 76}, 3814 (1995).
\bibitem{M69}
K. Maki in "Superconductivity", vol. 2, R. D. Parks ed.,
M. Dekker (New York, 1969).
\bibitem{U70}
K. D. Usadel, Phys. Rev. Lett. {\bf 25}, 507 (1970).
\bibitem{RS86}
J. Rammer and H. Smith, Rev. Mod. Phys. {\bf 58}, 323 (1986).
\bibitem{BW99}
W. Belzig, F. K. Wilhelm, C. Bruder, G. Sch\"on and A. Zaikin,
Superlattices and Microstructures, in press.
\bibitem{NS96}
Yu. V. Nazarov and T. S. Stoof, Phys. Rev. Lett. {\bf 76},
823 (1996).
\bibitem{Nota1}
This result (the value of the gap and the form 
of the energy spectrum, Figs.2,3 and 4) is insensitive to the
 morphology at atomic dimensions
 of the center of the neck, where $R \ll \xi$, and which
 determines $N$ and $T_n$ (see discussion of Fig.1).
\end{references}
\end{document}